\documentclass[12pt,preprint]{aastex}

\slugcomment{Accepted for publication in Astrophysical Journal}

\shorttitle{Sequestration of noble gases by H$_3^+$}
\shortauthors{Mousis et al.}

\begin{document}


\title{Sequestration of noble gases by H$_3^+$ in protoplanetary disks and outer solar system composition\\
}


\author{O. Mousis}
\affil{Institut UTINAM, CNRS-UMR 6213, Observatoire de Besan\c{c}on, BP 1615, 25010 Besan\c{c}on Cedex, France}
\email{olivier.mousis@obs-besancon.fr}

\author{F. Pauzat, Y. Ellinger}
\affil{Laboratoire de Chimie Th{\' e}orique (LCT/LETMEX), CNRS-UMR 7616, Universit{\' e} Pierre et Marie Curie, 4, Place Jussieu, 75252, Paris cedex 05 - France}

\and
\author{C. Ceccarelli}
\affil{Laboratoire d'Astrophysique, Observatoire de Grenoble, BP53, 38041 Grenoble Cedex 09, France\\
}




\begin{abstract}
  We study the efficiency of the noble gases sequestration by the ion H$_3^+$ in the form of XH$_3^+$ complexes (with X = argon, krypton or xenon) in gas phase conditions similar to those encountered during the cooling of protoplanetary disks, at the epoch of icy planetesimals formation. We show that XH$_3^+$ complexes form very stable structures in the gas phase and that their binding energies are much higher than those involved in the structures of X-H$_2$O hydrates or pure X-X condensates. This implies that, in presence of H$_3^+$ ions, argon, krypton or xenon are likely to remain sequestrated in the form of XH$_3^+$ complexes embedded in the gas phase rather than forming ices during the cooling of protoplanetary disks. The amount of the deficiency depends on how much H$_3^+$ is available and efficient in capturing noble gases. In the dense gas of the mid-plane of solar nebula, H$_3^+$ is formed by the ionization of H$_2$ from energetic particles, as those in cosmic rays or those ejected by the young Sun. Even using the largest estimate of the cosmic rays ionization rate, we compute that the H$_3^+$ abundance is two and three orders of magnitude lower than the xenon  and krypton abundance, respectively. Estimating the ionization induced by the young Sun, on the other hand, is very uncertain but leaves the possibility to have enough H$_3^+$ to make krypton and xenon trapping efficent. Finally, additional source of H$_3^+$ formation may be provided by the presence of a nearby supernova, as discussed in the literature. All together may cause a deficiency of Kr, Xe and, to a lower extent, of Ar in the forming icy planetesimals and, consequently, in the larger bodies accreted from them.  Recent solar system observations show a deficiency of Ar, and, even more, of Kr and Xe in Titan and in comets. In this article, we consider the possibility that this deficiency is caused by the afore-mentioned process, namely trapping of those noble gases by H$_3^+$ ions in the solar nebula. Assuming that this is the case, we show that this sequestration mechanism can be invoked in the formation zone of Titan's planetesimals in order to explain the clear deficiency of the satellite in noble gases revealed by the recent {\it Huygens} probe measurements. We also argue that comets formed from crystalline water ice in the outer solar nebula should be deficient in krypton and xenon, and to a lower extent, in argon, in agreement with some recent observations.
\end{abstract}


\keywords{Astrochemistry --- comets: general  --- planets and satellites: formation --- solar system: formation  --- planetary systems: protoplanetary disks}



\section{Introduction}

Noble gases are an important component of the atomic population in the
universe. Even if Helium is by far the most abundant, representing
about 10 \% of Hydrogen, the abundances of the heavier noble gases are
not negligible compared with other elements such as carbon or
oxygen. Moreover, noble gases seem to be widespread in the bodies of
the solar system located at heliocentric distances going up to $\sim$5 AU. Indeed, these compounds have been measured {\it in situ} in the
atmospheres of the Earth, Mars and Venus, as well as in meteorites
(Owen et al. 1992). The abundances of argon, krypton and xenon were
also measured oversolar by the {\it Galileo} probe in the atmosphere
of Jupiter (Owen et al. 1999).

However, even if there is a deficiency of accurate measurements, the
situation is likely to be different in bodies presumably formed at
further heliocentric distances. Indeed, an unexpected feature of the
atmosphere of Titan is that no noble gases other than argon have been
detected by the Gas Chromatograph Mass Spectrometer (GCMS) aboard the
{\it Huygens} probe during its descent on January 14, 2005. The
detected argon includes primordial ${}^{36}$Ar (the main isotope) and
the radiogenic isotope ${}^{40}$Ar, which is a decay product of
${}^{40}$K (Niemann et al. 2005). The other primordial noble gases
$^{38}$Ar, Kr and Xe were not detected by the GCMS
instrument. Besides, ${}^{36}$Ar/${}^{14}$N is about six orders of
magnitude lower than the solar value, indicating that the amount of
trapped ${}^{36}$Ar is poor within Titan (Niemann et al. 2005). The
apparent deficiency of noble gases in Titan is surprising since this
satellite was expected to be made from the same building blocks than
those accreted by Saturn (Mousis et al. 2002) and sharing a
composition similar to that of the noble gases-rich planetesimals that
took part to the formation of Jupiter, if the gas phase conditions
were homogeneous in the solar nebula (Alibert et al. 2005).

Regarding comets, there is no firm detection of noble gases,
in part due to the difficulty of the observations\footnote{The
    resonance transitions of the noble gases lie in the
    far-ultraviolet spectral region ($\lambda$ $\le$ 1200 \AA).} which
  need out-of-the-atmosphere instruments. Stern et al. (2000)
claimed the detection of argon in Comet C/1995 O1 Hale-Bopp, but the
signal to noise ratio is very low (Weaver et al. 2002; Iro et
al. 2003). On the contrary, Weaver et al. (2002) did not detect
$^{36}$Ar in Comets C/1999 T1 (McNaught-Hartley), C/2001 A2 (LINEAR)
and C/2000 WM1 (LINEAR), using the Far Ultraviolet Spectroscopic
Explorer (FUSE). The upper limits they obtained for C/2001 A2 and
C/2000 WM1 correspond to Ar/O ratios less than 10\% of the solar
value\footnote{The Ar/O upper limit for C/1999 T1 is essentially the
  solar value. However, the measurement was much less significant in
  that case, due to the fact that most of the data were accumulated
  during the daytime portion of the FUSE orbit and suffered some
  contamination from terrestrial day-glow emissions (Weaver et
  al. 2002).}. In summary, although the apparent deficiency of
noble gases in comets needs confirmation, current observations
  point to some process that affected their trapping capabilities
during their formation or evolution.

In this article, we explore the possibility that the observed deficit
in the afore-mentioned bodies is due to the sequestration of the noble
gases by H$_3^+$ during the early stages of the Solar nebula. The idea
is that, at the epoch of formation of the icy planetesimals by
accretion of the dust grains in the outer solar nebula, a significant
fraction of noble gases were sequestrated by the H$_3^+$ ion in the
gas phase.  As a result, they were not incorporated in the forming
ices during the cooling of the solar nebula, giving rise to noble gas
deprived planetesimals. From calculations of the H$_3^+$ abundance in young proto-planetary disks, we suggest that the proposed sequestration scenario might explain the observed deficit of Kr, Xe and, to a small extent, that of Ar.

For many years after they were discovered, it was believed that noble gases were chemically inert and unable to participate in molecular structures. Counterexamples can be found in X-halides species 
(Christe 2001) and in a novel class of noble gas insertion compounds recently discovered that is unlikely to form in space (Gerber 2004). Besides, van der Waals complexes of noble gases were also characterized but their stability due to neutral - neutral dispersion forces is very weak and can hardly resist turbulence.

However, this statement may
not be relevant when positive ions are involved, especially in the
case of the interstellar medium (ISM) where positive ions do exist in
space as free flyers. These species are mainly protonated species
formed by proton transfer from H$_3^+$ that plays in the gas phase the
role taken by H$_3$O$^+$ in the aqueous phase on Earth.

The fact that H$_3^+$ may be a possible partner for accreting noble
gases is a reasonable assumption since it is the simplest ion derived
from the most abundant molecule in the universe, H$_2$. This
assumption is comforted by recent detection of this molecular ion in a
large variety of environments: in star forming regions (Geballe \& Oka
1996), in diffuse interstellar media (McCall et al. 2003), at the
poles of Jupiter (Maillard et al., 1990) and of the other giant
planets (Miller et al. 2000). In particular, and relevant to the
subject of this article, it has also been detected under its
deuterated form H$_2$D$^+$ in protoplanetary disks (Ceccarelli et
al. 2004). All these observations show that H$_3^+$ and its deuterated
forms (H$_2$D$^+$, HD$_2^+$ and D$_3^+$) are likely the most abundant
positive charge carriers in cold and dense gas, as it is the case in
protoplanetary disks.

The formation of molecular complexes of composition (H$_2$)$_n$H$_3^+$
is well known in the laboratory (Clampitt \& Gowland, 1969; van
Deursen \& Reuss 1973; van Lumig \& Reuss 1978;
Hiraoka,1987). Concerning noble gases, Hiraoka \& Mori (1989) obtained
thermochemical data on ArH$_3^+$ and an estimation of the binding
energy of NeH$_3^+$. Apart from the latter examples, very few is known
on these complexes, mainly as a reason of extreme difficulties in
setting the proper experiments. In such conditions, numerical
simulations using Òstate of the artÓ methods of quantum chemistry can
be a valuable alternative.

In the following of this interdisciplinary report (Sec. 2), we show
that molecular complexes with H$_3^+$ are stable enough to trap noble
gases and serve as a means for maintaining these species in the gas
phase. We assume here a two-body reaction, where H$_3^+$ and the
  noble gas combine via radiative association. Since we are dealing
with protoplanetary disks conditions, the stability of these complexes
in the presence of H$_2$O has also to be addressed. Model calculations
lead to the conclusion that the transfer of the noble gas from
XH$_3^+$ compounds (where X is Ne, Ar, Kr or Xe) to the X-H$_2$O
complex or to the pure X-X condensate is unlikely. In Sec. 3, we
  discuss a possible implication of the production of such stable
  complexes formed by H$_3^+$ and noble gases for the composition of
  bodies formed in the outer solar nebula. Since H$_3^+$ results from
  H$_2$ ionization by cosmic rays, it is expected to be among the most
  abundant ions in the outer part of protoplanetary disks, including
  the outer solar nebula. Hence, H$_3^+$ could play a major role in
the gas phase chemistry leading to the formation of icy planetesimals
in the outer regions of the solar nebula (Pauzat et al. 2006). We
conclude in Sec. 4. Finally, an appendix section is devoted to the
description of the theoretical spectra that would allow the
identification of XH$_3^+$ complexes in laboratory.

\section{Chemical aspect of the complexation of noble gases by H$_3^+$}
\label{calculs} 

The quantum chemistry study presented below is a follow-up of a previous work that considered Argon as a case study in a systematic investigation involving possible multiple complexation of H$_3^+$ by this noble gas (Pauzat \& Ellinger 2005). 

Elaborate ab-initio methods of Coupled Cluster type (CCSD, CCSD(T)) and Density Functional Theory (DFT) methods using the BH\&HLYP
functionals were employed. For X=Ne, Ar and Kr, the basis sets used are extended basis with high flexibility in the valence shell (cc-pVTZ to Aug-cc-pVQZ); the same basis sets, associated with relativistic pseudo potentials, were used for Xe.

The details of the calculations, well beyond the scope of this report, can be found in the afore-mentioned paper and for specific technical points (correction for Basis Set Superposition Error, role of the extension of the basis sets and levels of excitations considered in the treatment of the electronic correlation) in a specialized paper (Pauzat \& Ellinger 2007). 

Consequently, only the results the most pertinent to the present study, i.e. those obtained at the post Hartree-Fock CCSD-T level of theory using the cc-pVQZ extended basis set are reported here. The outline of the computational procedure can be briefly summarized as follows:

\begin{description}
\item[(i)] optimization of all stationary points (minima and
  transition states) and verification of their stability conditions
  through vibrational analysis.
\item[(ii)] reconstruction of the potential energy surface around the
  minima by means of the counterpoise method (Boys \& Bernardi, 1970)
  in order to correct for the artefact known as basis set
  superposition error (BSSE) and notably large for weakly bound
  complexes.
\item[(iii)]	calculation of the rotational constants B$_e$.
\item[(iv)] calculation of the infrared spectra within the harmonic
  approximation and scaling of the frequencies relative to the
  hydrogenated fragment to account for anharmonic effects and
  incompleteness in the basis set. No ro-vibrational corrections are
  evaluated.
\end{description}
All calculations were performed using methods and basis sets as implemented in the Gaussian package (Gaussian, 2004).

\subsection{Structure and stability of XH$_3^+$ complexes}
\begin{figure}
\centerline{\includegraphics[width=10cm,angle=0]{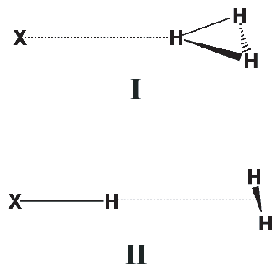}}
\caption{Stable structures of  XH$_3^+$.}
\label{Geom}
\end{figure}

The major result of this theoretical study is that all XH$_3^+$ are
stable with respect to dissociation.  An important point to be
mentioned is the gradual evolution of the complex (Fig. \ref{Geom})
that changes from a noble gas atom interacting with an almost
unchanged H$_3^+$ ion (I) to a protonated noble gas slightly elongated
by interaction with a neutral H$_2$ (II). It is illustrated by the
self explanatory Table \ref{Geometries} in which the geometries of the
complexes together with those of the constituting fragments are
reported.

From an energetic point of view, two dissociation mechanisms are
relevant to this study, namely:

\begin{equation}
XH_3^+ \rightarrow X + H_3^+
\end{equation}
\begin{equation}
XH_3^+ \rightarrow XH^+ + H_2
\end{equation}

\noindent The values listed in Table \ref{binding} show that all complexes are
stable with respect to either dissociation routes, which suggests that
H$_3^+$ can possibly act as a sink for noble gases. Looking in more
details at dissociation energies $D_e$, it can be seen that the family
of the noble gases splits in two. The first two complexes with Ne and
Ar dissociate according to reaction (1) while the last two complexes
with Kr and Xe follow reaction (2). It is a direct consequence of the
relative order of the proton affinity (PA) of the noble gases with
respect to molecular hydrogen (Dixon \& Lias 1987; Klein \& Rosmus
1984): PA$_{\rm (Ne)}$ = 49.4 $<$ PA$_{\rm (Ar)}$ = 90.6 $<$ PA$_{\rm
  (H_2)}$ = 101.4 $<$ PA$_{\rm (Kr)}$ = 108.1 $<$ PA$_{\rm (Xe)}$ =
124.8 in kcal/mol. In view of the available experiments, comparison of
binding energies can essentially be done for argon. From spectroscopic
studies, Bogey et al. (1988) were able to estimate $D_e$ at 6.6
kcal/mol or 9.9 kcal/mol according to the shape of the energy
potential used to analyze the data while Hiraoka \& Mori (1989)
proposed a value of $D_e$ = 6.69 kcal/mol for ArH$_3^+$ using
thermochemistry plots deduced from the mass spectrometry
experiments. For neon, the same technique provided an estimation of
$D_e$ at $\sim$0.4 kcal/mol. The theoretical values of 7.27 kcal/mol
and 0.98 kcal/mol obtained at the CCSD(T)/cc-pVQZ level for ArH$_3^+$
and NeH$_3^+$ respectively are both $\sim$0.6 kcal/mol higher than
those estimated from mass spectroscopy. The predictions for the other
dissociation energies of Table \ref{binding} should be equally accurate.

A point also worth mentioning is that H$_3^+$ is able to trap more than one atom of noble gas, although the binding energies of the second, $D_e(2)$ and third $D_e(3)$ atoms are weaker than that of the first one. In the case of argon, the values deduced by Hiraoka \& Mori (1989) from the thermochemistry plots ($D_e(2)$=4.6 and $D_e(3)$=4.3 kcal/mol) and confirmed by Pauzat \& Ellinger (2005) at the CCSD/cc-pVTZ level of theory ($D_e(2)$=3.9 and $D_e(3)$=3.5 kcal/mol) are still large enough to support the hypothesis of multiple trapping.

\subsection{Stability of XH$_3^+$ complexes in presence of ices}
\label{stab}

In this Section, we explore the stability of XH$_3^+$ complexes in
presence of different kind of ices (noble gases trapped in clathrate
hydrates structures or in the form of pure condensates) that can be found at low
temperature in protoplanetary disks (see Sect. \ref{predictions}). It
is worthwhile to mention that, for other volatiles like, for example
CO and CO$_2$, H$_3^+$ will preferably react to form the already known
protonated derivatives HCO$^+$ (Buhl \& Snyder 1970) and HOCO$^+$
(Thaddeus et al. 1981) because of the higher proton affinity of the
neutral parents compared to H$_2$.

The case of H$_2$O is more relevant to this study in view of the
general clathrates formula: (H$_2$O)$_n$. At this point, it should be
first noted that the binding energy of noble gases to the water
molecule is about an order of magnitude smaller than their binding
energy to H$_3^+$. The case of neon, however, should be considered
more carefully since the NeH$_3^+$ complex may be too weakly bound to
resist turbulence even at temperatures below 30 K.

Furthermore, the high proton affinity of H$_2$O compared to H$_2$
($\sim$170 compared to $\sim$101.4 kcal/mol) suggests that XH$_3^+$
and H$_2$O may not exist together: XH$_3^+$ could react with H$_2$O
giving H$_3$O$^+$ + H$_2$ + X in an exothermic process (between $\sim$50 and $\sim$70 kcal/mol for Ne to Xe). Such an energy released in
the clathrate hydrate is an order of magnitude larger than the 5.02
kcal/mol binding energy between two water molecules (Tschumper et
al. 2002) and about 5 times larger than the cohesion energy of
cristalline ice (Casassa et al. 2005). One can assume in these
conditions that XH$_3^+$ cannot be sequestrated in clathrates as
such. Its destruction, could even lead to their erosion through the
process detailed above (this phenomenon is currently under
investigation).

A second aspect to consider is the cohesion of pure noble gas solids ;
the values (in kcal/mol) known for cfc crystals are 0.46 (Ne), 1.84
(Ar), 2.67 (Kr), 3.91 (Xe). Compared to the binding energies of X to
H$_3^+$ reported in Table 2, they are 4 to 6 times smaller, implying
the idea that noble gases will remain under the form of XH$_3^+$
complexes. Another point to be mentioned at this level is the fact
that the binding energy of XH$_3^+$ is also more than an order of
magnitude larger than that of any XX noble gas dimer.

\section{Implications for the outer solar system composition}
\label{implications}
In this section, we explore the possibility that the mechanism
described previously, namely trapping of noble gases by the H$_3^+$
ions, is responsible for the deficiency of Ar, Xe and Kr observed in
Titan and possibly in comets, as suggested in the Introduction. The
idea is that ices entering the "vaporization line" at about 30 AU
sublimate in the early and hot nebula and later form clathrate
hydrates deficient in noble gases, these latters being sequestrated by
the H$_3^+$ ion at higher heliocentric distances than $\sim$5 AU (the
current Jupiter orbit, neglecting the possibility of migration during
its formation in the nebula). A critical point is, in this picture,
the H$_3^+$ abundance across the inner 30 AU of the solar nebula at
the time of the planetesimals formation. We address these two aspects
in the next two paragraphs. The last two paragraphs discuss the
possible consequences when the theory is applied to Titan and the
comets' formation respectively.

\subsection{Formation of planetesimals and the influence of H$_3^+$ trapping}
\label{predictions}

Formation scenarios of the protoplanetary nebula invoke two main
reservoirs of ices that took part in the production of icy
planetesimals. One reservoir, located within 30 AU of the Sun,
contains ices (mostly water ice) originating from the Interstellar
Medium (ISM) which, due to their proximity to the Sun, are initially
vaporized (Chick \& Cassen 1997). With time, the decrease of
temperature and pressure conditions allowed the water in this
reservoir to condense at $\sim$150 K in the form of microscopic
crystalline ice (Kouchi et al. 1994). It is then considered that most
of other volatile species were trapped as clathrate hydrates, and
eventually pure condensates\footnote{It is usually assumed that the
  local abundance of water is high enough to trap all the major
  volatiles in the form of clathrate hydrates in the outer solar
  nebula (Alibert et al. 2005; Mousis et al. 2006). In the contrary
  case, remaining volatiles that have not been trapped by water
  form pure condensates at lower temperatures in the nebula ($\sim$20-30 K).}, within 30 AU in the outer solar nebula (Mousis et
al. 2000). The other reservoir, located at larger heliocentric
distances, is composed of ices originating from ISM that did not
vaporize when entering into the disk. In this reservoir, water ice was
essentially in the amorphous form and the other volatiles remained
trapped in the amorphous matrix (Owen et al. 1999; Notesco \& Bar-Nun
2005). Consequently, icy planetesimals formed interior to 30 AU mainly
agglomerated from clathrate hydrates while, in contrast, those
produced at higher heliocentric distances (i.e. in the cold outer part
of the solar nebula) are expected to be formed from primordial
amorphous ice originating from ISM and containing noble gases trapped
in the amorphous water ice matrix.

\begin{figure}
\centerline{\includegraphics[width=11cm,angle=-90]{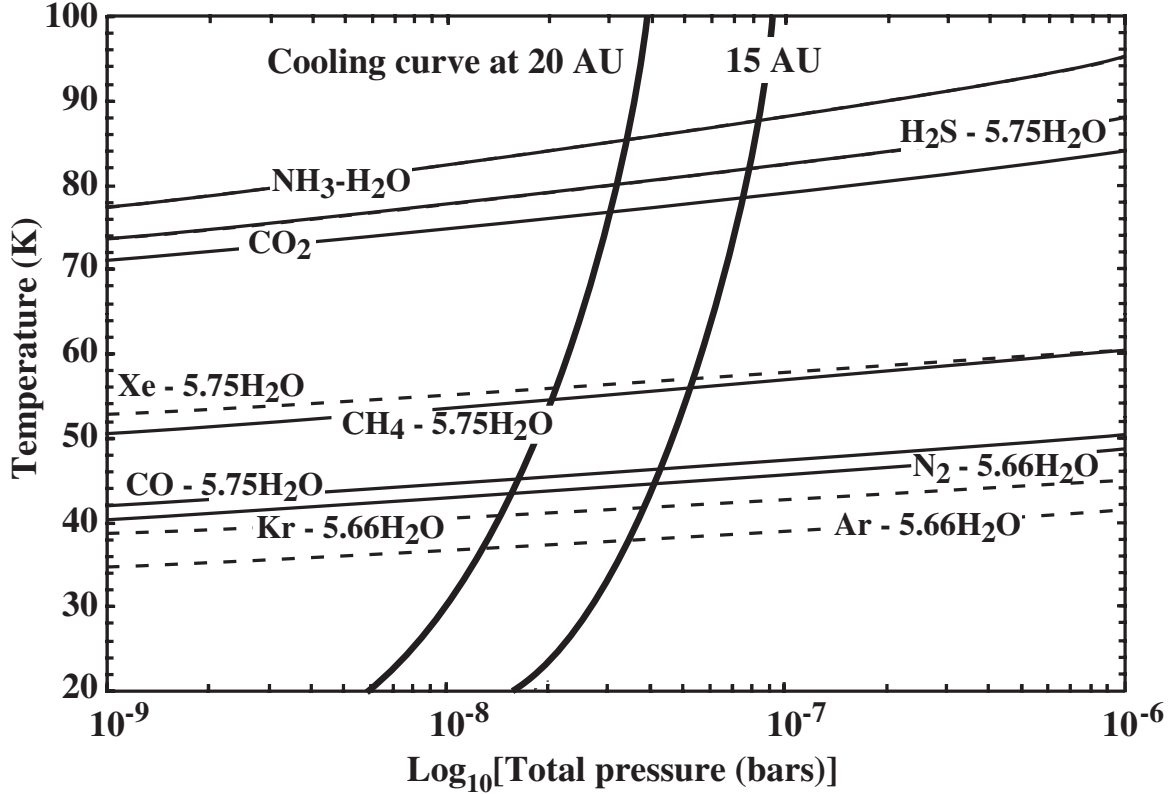}}
\caption{Stability curves of CO$_2$ pure condensate, NH$_3$ hydrate and clathrate hydrates of H$_2$S, CH$_4$, CO, N$_2$, Xe, Kr and Ar in the gas phase conditions of the solar nebula, and evolutionary tracks of the nebula at 15 and 20 AU. Abundances of various elements are solar, with CO$_2$:CO:CH$_4$ = 30:10:1 and N$_2$:NH$_3$ = 1 as molar mixing ratios in the gas phase. Species remain in the vapor phase as long as they stay in the domains located above the stability curves. Dashed lines correspond to the stability curves of clathrate hydrates of noble gases which form if one ignores the presence of H$_3^+$ in the vapor phase. If the sequestration of noble gases by H$_3^+$ is efficient, they remain in the solar nebula gas phase in the form of XH$_3^+$ complexes, even when the temperature and pressure conditions allow their trapping as clathrate hydrates (see text).}
\label{structures}
\end{figure}

In order to quantify this effect, we computed the ices abundance in
the cooling solar nebula, following the model described in Mousis et
al. (2006), with and without the trapping of noble gases by H$_3^+$.
The composition of ices formed within 30 AU in the solar nebula is
determined using the clathrate hydrate trapping theory, as presented
e.g. in Mousis et al. (2006). With time, the decrease of temperature
and pressure conditions in the solar nebula leads to the trapping of
volatile species as hydrates or clathrate hydrates, as illustrated in
Fig. \ref{structures}. The trapping of volatiles is calculated using
the stability curves of hydrates or clathrate hydrates and the
evolutionary tracks of temperature and pressure (hereafter cooling
curves) at given heliocentric distances, taken from a solar nebula
model (Alibert et al. 2005). The cooling curves intercept the
stability curves of the different ices at some given temperature and
pressure conditions. For each considered ice, the domain of stability
is the region located below its corresponding stability curve.  As a
result, in gas phase conditions where the presence of H$_3^+$ is
neglected, all major species\footnote{Except the unique cases of
  CO$_2$ and Ne. CO$_2$ crystallizes as a pure condensate prior to
  being trapped by water to form a clathrate hydrate in the solar
  nebula (Alibert et al. 2005; Mousis et al. 2006) and Ne is a poor
  clathrate hydrate former (Lunine \& Stevenson 1985).}, including Ar,
Kr and Xe, are presumed to be trapped under the form of hydrates or
clathrate hydrates during the cooling of the solar nebula (provided
the abundance of water is large enough).

On the other hand, assuming that Ar, Kr and Xe have been sequestrated
by H$_3^+$ in the form of XH$_3^+$ complexes in the gas phase, and
following the considerations developed in Sec. 2.2, they are expected
to have been preserved from the trapping as clathrate hydrates by
crystalline water at low temperature and pressure conditions in the
nebula. As a result, icy planetesimals formed in presence of a large
concentration of H$_3^+$ in the gas phase may not contain Ar, Kr and
Xe albeit they trapped most of the other major volatile species.

\subsection{The H$_3^+$ abundance in the inner 30 AU}
\label{inner} 
The ionization structure of the inner 30 AU proto-planetary disks, which are supposed to be a good description of the solar nebula at the time of the planetesimal formation, have been computed by a few authors, giving somewhat different results (see e.g. the discussion in Markwick et al. 2002). The most recent computations by Semenov et al. (2004) predict ionization fractions between $1\times10^{-15}$ and $1\times10^{-10}$ in the plane of the disk, where the only source of ionization are the cosmic rays. In this region, the positive charge carrier is HCO$^+$, and H$_3^+$ is somewhat lower in abundance. Although the overall ionization degree is a rather complex problem to solve, the abundance of H$_3^+$ in the inner 30 AU region is relatively simple to compute, {\it in principle}. In fact, H$_3^+$ is formed by the ionization of H$_2$ from energetic ($\leq 100$ MeV) particles, like those in the cosmic rays or those emitted by the young Sun, at a rate $\zeta_{ep}$, and destroyed by the reactions with the most abundant molecules, namely CO and H$_2$O, at a rate k$_{dCO}$ and k$_{dH_2O}$, respectively. At steady state, the formation and destruction rates balance, and this provides a relation between the H$_3^+$ abundance ($x_{H_3^+}$) and the CO and H$_2$O abundances ($x_{CO}$ and $x_{H_2O}$) with respect to H$_2$, the H$_2$ density $n_{H_2}$, and the H$_2$ ionization rate $\zeta_{ep}$, as follows:
\begin{equation}\label{eqh3+}
  x_{H_3^+} = \frac{\zeta_{ep}}{n_{H_2} \left(\kappa_{dCO}~x_{CO} + \kappa_{dH_2O}~x_{H_2O}\right)} 
\end{equation} 
\noindent
Note that in the inner 30 AU, the temperature in the mid-plane is considered larger than the temperature required for the condensation of ices so that the abundances of CO and H$_2$O are set equal to $1\times10^{-4}$, namely approximatively their gas phase abundances in the solar nebula (Mousis et al. 2006). The reaction rates $k_{dCO}$ and $k_{dH_2O}$ are respectively $1.7\times10^{-9}$ and $5.9\times10^{-9}$ cm$^3$s$^{-1}$ (from the UMIST database {\it http:www.rate99.co.uk}).

Equation \ref{eqh3+} shows that the H$_3^+$ abundance is inversely proportional to the H$_2$ density and linearly depends on the ionization rate $\zeta_{ep}$. We emphasize that Eq. \ref{eqh3+} only applies to the disk mid-plane where the electronic abundance is less than $\sim \mathrm{5\times10^{-7}}$, otherwise the H$_3^+$ recombination would be a major route of H$_3^+$ destruction, lowering the H$_3^+$ abundance. As mentioned above, there are at least two known sources of H$_3^+$ formation from the H$_2$ ionization: the cosmic rays and the energetic particles emitted from the young Sun. A third source could be a nearby supernova exploded during the Solar System formation (e.g. Ouellette et al. 2007 for a recent discussion of this hypothesis). Note that X-rays, although copiously
emitted from the young Sun, do not significantly affect the ionization in the mid-plane, as they cannot penetrate so deeply (e.g. Glassgold et al. 1997). In the following, we will discuss the first three sources of H$_3^+$ formation.

\begin{figure}
\centerline{\includegraphics[width=16cm,angle=0]{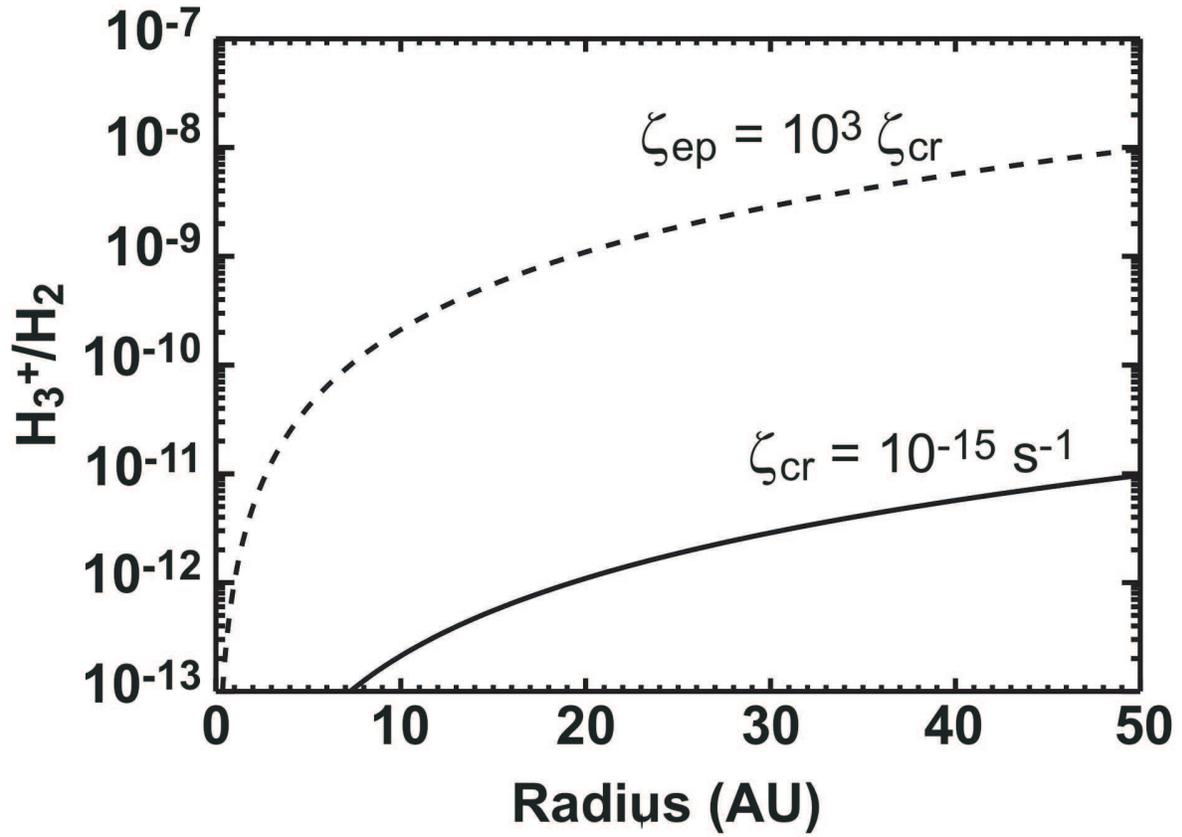}}
\caption{H$_3^+$ abundance in the disk midplane as a function of the distance from the center in the case of a cosmic rays ionization rate of 1 time the standard value and in the case of an enhancement by a factor of 1000 (see text).}
\label{ioniz}
\end{figure}

The cosmic rays ionization rate has been estimated in different regions, giving somewhat different results. A lower limit has been estimated to be around few times $10^{-17}$ s$^{-1}$ (e.g. Webber 1998), whereas the upper limit is around $1\times10^{-15}$ s$^{-1}$ ( Payne et al. 1984). Here we adopt the high value to consider the most optimistic case for the H$_3^+$ abundance. Figure \ref{ioniz} shows the computed H$_3^+$ abundance with respect to H$_2$ in the disk mid-plane as a function of the distance from the center for such high value. In these computations, we have used the structure of a 0.001 M$_\odot$ disk surrounding a 0.5 M$_\odot$ star with a luminosity of 0.5 L$_\odot$ and a temperature T$_{eff}$ of 3630 K (namely a typical T Tau star, believed to be a good representative of the Sun in the early phases of the Solar System formation). The average grain size is assumed to be 1 $\mu$m, namely dust started to coagulate on the disk midplane. For reference, such a disk has a H$_2$ density of $\sim 10^{9}$ cm$^{-3}$ at around 30 AU and its physical structure has been reported in Ceccarelli \& Dominik (2005).  Figure  \ref{ioniz} shows that the H$_3^+$ production by the cosmic
rays only is (at least) 3 orders of magnitude lower than the Kr solar abundance reported in Table \ref{gas}.

That other sources of energetic particles existed in the early phase of the Solar System formation is testified by the presence of radionuclides with half-lives less than 1 Myr in the calcium-aluminum-rich
inclusions (CAIs) in meteorites (see for example the recent review by McKeegan \& Davis 2005). This includes light atoms, like $^{10}$Be and heavy atoms, like $^{60}$Fe. These two extremes, $^{10}$Be and $^{60}$Fe, imply two different origins: an {\it inner} and {\it external} production of energetic particles, respectively. By the {\it inner} origin is meant irradiation of energetic particles from the young Sun (e.g. Lee et al. 1998; Gounelle et al. 2001, 2006), whereas by {\it external} origin is meant the presence of a nearby supernova during the Solar System formation (e.g. Busso et al. 2003; Tachibana \& Huss 2003). The other short-lived radionuclides (e.g. $^{26}$Al, $^{36}$Ca and $^{41}$Ca) can be produced by models considering both origins with a different ``mix'' (see for example the discussion in Gounelle et al. 2006). The analysis of the observed X-rays luminosities and spectra in Young Stellar Objects implies a flux of energetic particles about $1\times10^{5}$ times that of the Sun today (Feigelson et al. 2002), with flares exceeding $1\times10^{3}$ times more (Leya et al. 2003). However, the exact value of the flux of energetic particles {\it through the disk} as well as its spectrum and the influence on the disk ionization and chemical composition are highly uncertain. For this reason, instead of trying to model the two scenarios, the inner and external irradiation of energetic particles respectively, we computed the minimum enhancement of energetic particles flux with respect to the cosmic rays in order to make the H$_3^+$ abundance larger than the krypton abundance. This is the minimum to make H$_3^+$ able to trap krypton, assuming that the process has an efficiency close to unity. The results of these computations are shown in Fig. \ref{ioniz}: the flux of energetic particles has to be at least 3 orders of magnitude larger than the cosmic rays flux.  If these conditions apply to the young Sun -which may not be unreasonable- trapping of krypton (and, even more xenon) by H$_3^+$ might be an efficient mechanism able to explain the deficiency of these two elements, and to a lower extent that of Ar\footnote{ It is important to note that even if the H$_3^+$ abundance is postulated high enough in the disk to allow an efficient trapping of argon (at least $10^{-6}$ times the H$_2$ abundance), then the dissociative reaction H$_3^+$ + e$^-$ $\rightarrow$ H$_2$ + H, with a reaction rate of $\sim$$10^{-7}$ cm$^3$ s$^{-1}$, becomes competitive. This can then further deplete H$_3^+$ in the disk and lead to the break-up of Ar-H$_3^+$ if its recombination rate is similar.} in some bodies of the outer Solar System. In the following, we will assume that this is the case and explore in more detail this possibility.

\subsection{Interpreting the noble gases deficiency in Titan}
\label{Titan}

Several different scenarios interpreting the deficiency of Titan's atmosphere in primordial noble gases have been recently proposed. These scenarios can be regrouped in two different approaches. The first  category of scenarios postulates that Titan initially incorporated primordial noble gases during its formation but that subsequent processes prevented them to be detected in its atmosphere. For example, Jacovi et al. (2005) suggested that the aerosols observed in Titan's atmosphere may have cleared its content in noble gases, assuming they were produced from the aggregation of polymers. Alternatively, Osegovic \& Max (2005) proposed that clathrate hydrates formed from hydrocarbons and nitrogen on the surface of Titan are possibly a sink for primordial Ar, Kr and Xe absent from its atmosphere. On the other hand, Thomas et al. (2007) reinvestigated the conclusions of Osegovic \& Max (2005) and demonstrated that Ar is a poor clathrate hydrate former on the surface of Titan, contrary to Kr and Xe. These authors then concluded that an alternative scenario must be proposed in order to explain the Ar deficiency in the atmosphere of Titan. The second category of scenarios, which is favored in this work, postulates that the nobles gases deficiency in Titan's atmosphere results from processes that occured either during its accretion or during the formation of its planetesimals. These scenarios present the advantage to explain in a self-consistent way the current characteristics of Titan's atmosphere and not only its apparent deficiency in noble gases. By this way, in order to interpret the observed properties of Titan's atmosphere\footnote{Although the thermochemical calculations predict that carbon essentially exists in the form of CO and CO$_2$ in planetesimals formed in the outer solar nebula (Mousis et al. 2006), the main carbon species detected in the atmosphere of Titan is CH$_4$, with a mole fraction of $\sim$5 \% close to the surface (Niemann et al. 2005). The atmosphere of Titan is dominated by N$_2$ and CH$_4$. CO is several orders of magnitude less abundant than CH$_4$ and the existing nitrogen is presumed to be the product of ammonia photolysis or shock chemistry (Atreya et al. 1978, McKay et al. 1988).}, and ignoring the possibility of formation of XH$_3^+$ complexes in the gas phase,
 Alibert \& Mousis (2007) proposed that its building blocks were produced in Saturn's feeding zone and/or in its surrounding subdisk. They then argued that, during their formation, these planetesimals would have incorporated all the main volatile compounds, including noble gases, by following the clathration sequence described in Fig. \ref {structures}. During their migration within the subnebula, the planetesimals that led to the formation of Titan would have suffered a partial vaporization, due to the higher temperature and pressure conditions encountered in the gas phase on their pathway, thus leading to their impoverishment in clathrate hydrates whose dissociation temperatures are the lowest. As a result, Titan would be formed from planetesimals deficient in CO and in noble gases, in agreement with its observed properties.

Following the scenario proposed by Alibert \& Mousis (2007), and similarly to Jupiter, the abundances of noble gases within Saturn's atmosphere should be oversolar since the giant planet accreted volatiles-rich planetesimals produced in its feeding zone. Table \ref{enrichissements} reproduces the enrichment in volatiles predicted in Saturn by Mousis et al. (2006) for a nominal solar nebula gas phase composition (CO$_2$:CO:CH$_4$ = 30:10:1 and N$_2$:NH$_3$ = 1), in agreement with the scenario of Alibert \& Mousis (2007), and assuming that the amount of available crystalline water was sufficient to trap all the main volatile species.

On the other hand, if the sequestration of noble gases by H$_3^+$ was efficient at the heliocentric distance of 10 AU in the solar nebula gas phase, Ar, Kr and Xe should be poorly trapped in icy planetesimals formed in this location. This hypothesis is fully compatible with the mechanism proposed by Alibert \& Mousis (2007), namely a partial vaporization of its building blocks during their migration within the subnebula, which is required to explain at least the deficiency in CO in its atmosphere. In order to discriminate which of the two processes was likely at 10 AU in the solar nebula (sequestration of noble gases by H$_3^+$ or their full clathration by crystalline water), we predict that abundances of Kr and Xe should remain solar within Saturn's atmosphere (see Table \ref{enrichissements}), if the formation of XH$_3^+$ complexes (with X = Kr or Xe) was efficient in the gas phase. Indeed, since these XH$_3^+$ complexes would stay coupled with the gas phase, their molar mixing ratios relative to hydrogen should still remain constant, in particular within Saturn's envelope which results from the accretion of gas and gas-coupled solids from the solar nebula.

The case of Ar is less straightforward since its solar abundance is at least $\sim$3 orders of magnitude higher than those of Kr and Xe (see Table \ref{gas}). Following our discussion in Sec. 3.2, requiring a H$_3^+$ abundance high enough to trap efficiently Ar in the solar nebula is probably unlikely. This is in agreement with the detection by {\it Huygens} of only ${}^{36}$Ar, among the primordial noble gases, in Titan. On the other hand, the Ar subsolar abundance measured in the satellite's atmosphere suggests that it has been only partially trapped in planetesimals produced in Saturn's feeding zone. As a result, we estimate that the corresponding Ar abundance within Saturn's atmosphere is enclosed by two extreme values: a solar abundance and the oversolar prediction made by Mousis et al. (2006) (see Table \ref{enrichissements}). The inner and upper values correspond respectively to the full sequestration of Ar by H$_3^+$ in the solar nebula gas phase and a full trapping of the noble gas in the form of clathrate hydrate in the forming planetesimals.

\subsection{Implications for comets}

 The origin of comets in the solar nebula is still poorly constrained: Edgeworth-Kuiper belt comets are presumed to be formed at a distance further to the Sun than that of Neptune prior its migration and the formation zone of Oort cloud comets varies, according to the different models, from near Jupiter and Saturn to heliocentric distances higher than 30 AU. As a function of the considered formation scenario (see Horner et al. 2007 for a comprehensive review), comets may have formed either from clathrate hydrates in the initially hot part of the solar nebula (where the local gas temperature was initially high enough to vaporize the entering ices) or from amorphous ice at heliocentric distances higher than 30 AU. This implies that the content of comets in noble gases may vary as a function of their formation region. From the scenario proposed in Sec. \ref{predictions} and following the argumentation developed in the case of Ar in Sec. \ref{Titan}, we propose that comets formed within the heliocentric distance of 30 AU should be depleted efficiently in Kr and Xe, and only partly in Ar, in agreement with the marginal detection of this species in Comet C/1995 O1 Hale-Bopp by Stern et al. (2000). On the other hand, comets formed at higher heliocentric distances from an initially amorphous ice mixture should contain Ar, Kr and Xe. In that context, subsequent measurements of noble gases abundances in comets should help to constrain the origin of the reservoirs of cometary bodies.

\section{Summary and conclusions}

In this work, we have studied the efficiency of the noble gases sequestration by the H$_3^+$ ion in the form of XH$_3^+$ complexes (with X = argon, krypton xenon or neon) in gas phase conditions similar to those encountered during the cooling of protoplanetary disks and, especially, at the epoch of the planetesimals formation. We have calculated that the binding energy of noble gases to the water molecule is $\sim$one order of magnitude lower than their binding energy to H$_3^+$, meaning that the XH$_3^+$ complex is more stable than a clathrate hydrate of species X. Similary, we have shown that the binding energy of XH$_3^+$ complexes is more than one order of magnitude higher than that of pure condensates of noble gases. This implies that, provided H$_3^+$ is abundant enough, once argon, krypton or xenon are sequestrated in the form of XH$_3^+$ complexes in the gas phase, they remain in this structure rather than forming ices during the cooling of protoplanetary disks. The
 case of neon should be considered more carefully since the NeH$_3^+$ complex may be too weakly bound to resist turbulence even at low temperatures in the nebula.

On the other hand, 
we have calculated that the abundance of H$_3^+$ in the 3 to 30 AU mid-plane region of disks is about $1\times10^{-12}$ at most, implying a poor noble gases sequestration (the abondances of xenon, krypton and argon hold between $\sim$$10^{-10}$ and $10^{-6}$). However, we have stressed that there is evidence that the solar nebula has been exposed to other sources of energetic particles. This is testified by the presence of radionuclides with half-lives less than 1 Myr in the calcium-aluminum-rich inclusions in meteorites which could imply two different origins: an {\it inner} and {\it external} production of energetic particles, respectively. By the {\it inner} origin is meant irradiation of energetic particles from the young Sun, whereas by {\it external} origin is meant the presence of a nearby supernova during the Solar System formation. If this hypothesis is correct, one associated effect might be an augmentation of the cosmic ionization rate in the disk by several orders of magnitude. From these statements, given the noble gases abundances in the solar nebula, Kr and Xe could be relatively easily trapped, in regions where H$_3^+$ is at least $\sim$$1\times10^{-9}$ times the H$_2$. Argon, with its higher abundance, may be more difficult to trap and requires a minimum H$_3^+$ abundance of $\sim$$1\times10^{-6}$ times the H$_2$.

Taking into account these considerations, we have assumed that the H$_3^+$ abundance in the 10 -- 30 AU region of the solar nebula was indeed effectively large enough to make possible at least the efficient trapping of xenon and krypton, and limited trapping of argon. Therefore, icy planetesimals formed from ices crystallized within this distance range in the solar nebula should be impoverished in krypton and xenon. This scenario allowed us to propose that the apparent deficiency of Titan's atmosphere in noble gases can be explained if the satellite was formed from such solids. Since Titan is expected to be made from the same building blocks than those accreted by Saturn, the abundances of at least krypton and xenon should be solar in the giant planet. The case of argon is less straightforward since its isotope $^{36}$Ar is the only primordial noble gas detected (in subsolar abundance ratio) in Titan's atmosphere. This is in agreement with the idea that argon has been sequestrated only partially by H$_3^+$ in the solar nebula.

Moreover, the proposed scenario allowed us to argue that comets formed from crystalline water ice in the outer solar nebula should be deficient in krypton and xenon, and to a lower extent, in argon, in agreement with some recent observations. Following our scenario, we also predict that the abondances of Kr and Xe in Uranus and Neptune should be solar, provided that the two planets have accreted planetesimals agglomerated from ices crystallized in the solar nebula. A more direct way for testing this scenario would be to measure the abundances of noble gases in a significant sample of comets. Some measurements will be obtained from planned space missions towards Jupiter family comets, such as the {\it Rosetta} mission whose target is the comet 67P /Churyumov-Gerasimenko. However, the determination of their composition could be ambiguous because such comets may have been initially formed from amorphous ice originating from the presolar cloud and may not be relevant for the processes that occured between 10 and 30 AU in the solar nebula. We note that, in order to make our H$_3^+$ sequestration scenario compatible with the noble gases oversolar abundances measured within Jupiter atmosphere by the {\it Galileo} probe, it is necessary to assume that the ion abundance was very low in the inner solar nebula up to the formation distance of the giant planet at least. This hypothesis needs to be tested in a near future.
 
Finally, we note that, at temperatures greater than $\sim$150 K in the solar nebula, the capability of H$_3^+$ to trap the noble gases can be limited. Indeed, due to the different proton hopping reactions such as those with H$_2$O, the lifetime of the XH$_3^+$ complex is very short (of the order of 10$^4$ s) compared with the time taken to form ices in the disk (several hundreds of thousand of years). As a result, in these temperature conditions, the sequestration of the noble gases by H$_3^+$ is probably not stable for sufficiently long periods of time against their re-release by proton hopping destruction, for this to be truly effective. On the other hand, the lifetime of the XH$_3^+$ complex is expected to significantly increase once the condensation temperatures of the different species involved in its destruction are reached in the disk, provided that the noble gases still remain in the gas phase. For example, at temperatures lower than $\sim$150 K in the outer nebula, water ice is formed and the abundance of H$_2$O molecule converges towards zero in the gas phase. This implies that the destruction rate of XH$_3^+$ by H$_2$O becomes insignificant. In order to be validated, these considerations will require the development of a hybrid thermochemical and chemical-kinetic model of the protosolar nebula.

\acknowledgments

The calculations presented in this contribution were financed by the CINES national supercomputing
facility. We thank Steve Miller, John Black, Alexandre Faure, Thierry Montmerle whose helpful comments invited us to improve our manuscript. We acknowledge the support of the French Programme National de Plan{\'e}tologie.

\appendix

\section{Theoretical spectra}
Every astrochemical model is based on observations or non-observations of chemical species. In the present case the observations of H$_3^+$ and H$_2$D$^+$ provided the incentive for this study. Now, observations of XH$^+$ and XH$_3^+$ complexes would be of utmost interest. At that point, it should be recalled that identification of a molecule in space relies on a perfect match between the observed lines and the spectrum of the same species in the laboratory. However, it is something difficult to realize, especially for the type of unstable species we are dealing with. Generally indeed, unstable species are formed in the experiments together with a number of other molecules that may be isomers or even compounds with completely different chemical formulas.  It is the purpose of the numerical simulations reported below to propose theoretical spectra that are precise enough to make possible an interpretation of the laboratory experiments to come and extract the spectra of the tar
 get molecules from the mixture of experimental values for comparison with further spatial observations.

\subsection{Rotational constants}
ArD$_3^+$ is the only complex for which spectroscopic data are available (ArH$_3^+$ is poorly characterized by a few lines). One has to keep in mind that rotational constants are not direct products of the experiments; like $D_e$, they are extracted from the very complex spectra of the clusters by searching the best possible match between the measured spectra and the one deduced from phenomenological models. Three of those models, hereafter referred to as rigid (RM), semi-rigid (SRM) and flexible (FLM) models, were successively used in which the H$_3^+$ fragment was first constrained to a frozen equilateral triangle, then allowed to freely rotate and finally relax during rotation. Distances were then adjusted to reproduce the experimental data to the best possible.

The $A$ constant of ArH$_3^+$ was first estimated at $A$=1309.2 GHz by scaling of the ArD$_3^+$ value according to the H/D ratio of the atomic masses within the RM approximation (Bogey et al. 1987); $B$ = 30.9623 GHz and $C$ = 30.1378 GHz were determined subsequently. Spectroscopic data on mixed H/D species made it possible to implement the SRM and FLM models (Bogey et al. 1988). Adjusting the constants in the semi-rigid model gives the following set of experimental values:

	$A$ = 1310 GHz, 		$B$ = 30.811 GHz, 	$C$ = 30.155 GHz.

The flexible model leads to a second set of experimental values:

	$A$ = 1490 GHz, 		$B$ = 31.411 GHz, 	$C$ = 30.750 GHz.

It can be seen from Table \ref{constants1} that the rotational constant $A$ corresponding to rotation about the C$_2$ axis of ArH$_3^+$ deduced from the FLM description is within $\sim$2\% of the theoretical value. The present quantum simulation  shows that the RM and SRM approximations are clearly inadequate for determining the $A$ constant since the $\sim$12\% error in the RM and SRM descriptions is much larger than the maximum error expected for the type of CCSD(T) calculations performed here. In these conditions it is clear that the quality of the results of the quantum calculation is at least of equal quality to that of the so-called experimental values deduced from the various spectroscopic models. The values predicted for the cluster series reported in Table \ref{constants1} should be accurate enough to stimulate new laboratory experiments and possible observations in the future. Apart from XeH$_3^+$ whose dipole moment ($\mu$ =2.6 Debye) is the weakest of the series
 , all other complexes with dipole moments in the range of $\sim$6-8 Debye should be reasonable targets for detection by microwave spectroscopy.

\subsection{Infrared signatures}
The harmonic vibrational frequencies of the XH$_3^+$ (X=Ne-Xe) are reported in Table \ref{constants2} together with the corresponding absolute intensities. These best estimated frequencies were deduced
from the CCSD(T)/cc-pVQZ level of theory using an appropriate scaling procedure.

The point to consider is that the spectrum of any of these molecules is clearly separated into two domains. At high frequencies one has the internal vibrations of each fragment; at low frequencies one finds the inter fragment vibrations corresponding to the relative motions of the two interacting species. Whenever related experimental data are available, it is a well-admitted procedure to use the ratio of the
calculated to observed frequencies for all vibrations higher than 2000 cm$^{-1}$ as scaling factor to derive a best estimate of the frequencies to look at. In the present case, the frequencies of the individual fragments H$_3^+$ (Oka 1980; Ketterle et al. 1989) and H$_2$ (Stoicheff 1957; Huber \& Herzberg 1979) are well known.

The scaling factors deduced from these values are then applied to the corresponding frequencies in the complexes. Not scaling lower frequencies is entirely justified in view of the raw spectrum calculated for ArD$_3^+$ (Pauzat \& Ellinger 2005): $\nu_1$ = 2542; $\nu_2$ = 1643; $\nu_3$ = 1504; $\nu_4$ = 570; $\nu_5$ = 437; $\nu_6$ = 351 (values in cm$^{-1}$) for which the value obtained for the $\nu_6$ stretching frequency is in complete agreement with the 352 cm$^{-1}$ obtained for the only ÒexperimentalÓ frequency that could be estimated from the centrifugal distortion in the ArD$_3^+$ experiment (Bogey et al. 1988). The changes in the frequencies from the ion in isolation confirm the structural changes of the comlexes when going from Ne to Xe. The most intense bands should be good candidates for more studies in the laboratory in view of possible application in space.

\clearpage

\begin{table}[h]
\caption[]{Optimized geometries (\AA) of XH$_3^+$ (X = Ne to Xe) after full correction of BSSE together with those of the constituting fragments (experimental values in italic)}
\begin{center}
\begin{tabular}[]{clll}
\hline
\hline
\noalign{\smallskip}
 & X-H$_X$ & H$_X$-H & H-H \\
\noalign{\smallskip}
\hline
\noalign{\smallskip}
NeH$_3^+$ &    1.7939                                  &   0.8821    &    0.8607     \\
NeH$^+$      &    0.9886    (0.9912 $^a$)    &                    	                     \\
ArH$_3^+$   &    1.8306                                 &  0.9342	     &    0.8280      \\
ArH$^+$        &    1.2800   (1.2804 $^b$)    &  	    	     &  	 	  \\
KrH$_3^+$   &    1.7236				 &  1.0575	     &	  0.7954      \\
KrH$^+$        &    1.4155    (1.4212 $^c$)    &                   &                      \\
XeH$_3^+$  &    1.6451     			  &	1.5304 & 0.7581	\\
XeH$^+$       &  1.5909    (1.6028 $^d$)	\\
\hline
H$_3^+$	&	&  0.8737  & 0.8737    (0.8734 $^e$)   \\
H$_2$	&	&  & 0.7416    (0.7414 $^f$)   \\
\hline
\end{tabular}
\end{center}
\tablecomments{$^a$Ram et al. 1985; $^b$Johns 1984; $^c$Warner et al. 1984; $^d$Rogers et al. 1987; $^e$Cencek et al. 1998; $^f$Huber \& Herzberg 1979}
\label{Geometries}
\end{table}

\clearpage

\begin{table}[h]
\caption[]{Binding energies (kcal/mol) for XH$_3^+$ (X= Ne to Xe) including full corrections of BSSE and zero point vibration energies.}
\begin{center}
\begin{tabular}[]{ccccc}
\hline
\hline
\noalign{\smallskip}
 & Ne 	& Ar 		& Kr 		& Xe \\
\noalign{\smallskip}
\hline
\noalign{\smallskip}
XH$_3^+$ $\rightarrow$ X + H$_3^+$   	    &  0.98  		& 	 7.28 	&	13.05	&	23.22 \\
XH$_3^+$ $\rightarrow$ XH$^+$ + H$_2$  &  50.57 	&	16.00		& 	8.60		& 	2.92    \\
\hline
\end{tabular}
\end{center}
\label{binding}
\end{table}

\clearpage

\begin{table}[h]
\caption[]{Protosolar abundances (molar mixing ratio with respect to
H$_2$) of noble gases of interest (from Grevesse et al. 2005).}
\begin{center}
\begin{tabular}[]{lc}
\hline
\hline
Species & abundance \\
\hline
Ar & $3.02 \times 10^{-6}$ \\
Kr & $3.82 \times 10^{-9}$ \\
Xe  & $3.72 \times 10^{-10}$ \\
\noalign{\smallskip}
\hline
\end{tabular}
\end{center}
\label{gas}
\end{table}

\clearpage

\begin{table}[h]
\caption[]{Observed carbon enrichment in Saturn (Flasar et al. 2005), and calculated enrichments in volatiles assuming a full clathration of noble gases by crystalline water (Mousis et al. 2006) or their full sequestration in the form of XH$_3^+$ complexes in the solar nebula gas phase (this work). Enrichments are calculated using the nominal gas phase conditions given in Mousis et al. (2006) (CO$_2$:CO:CH$_4$ = 30:10:1 and N$_2$:NH$_3$ = 1) and are calibrated on the observed carbon enrichment.}
\begin{center}
\begin{tabular}{lccc}
\hline
\hline
\noalign{\smallskip}
Species &  Observed & Mousis et al. 2006 & This work  \\
\noalign{\smallskip}
\hline
\noalign{\smallskip}
    C		&	$8.8 \pm 1.7$		& 7.1		& 7.1 \\
    N   	& 					& 6.5 	& 6.5\\
    S   	&   					&  5.5 	&  5.5\\
    Ar  	&    					& 5.1 	& 1 to 5.1$^{\star}$\\
   Kr  		&    					& 5.6 	& 1 \\
   Xe 		&   					&  6.7 	& 1 \\
\hline
\end{tabular}
\end{center}
\tablecomments{$^{\star}$ As a function of the assumed argon trapping scenario (see text).}
\label{enrichissements}
\end{table}

\clearpage

\begin{table}[h]
\caption[]{Rotational constants (GHz) and dipole moment (Debye) of XH$_3^+$ (X=Ne to Xe).}
\begin{center}
\begin{tabular}[]{lcccc}
\hline
\hline
\noalign{\smallskip}
Complex & $A$ 	& $B$ 		& $C$ 		& $\mu$ \\
\noalign{\smallskip}
\hline
\noalign{\smallskip}
NeH$_3^+$	&	1353.9	&	35.147	&	34.258	&	7.9	\\
ArH$_3^+$	&	1463.0	&	30.607	&	29.979	&	7.4	\\
KrH$_3^+$	&	1585.1	&	29.502	&	28.963	&	6.5	\\
XeH$_3^+$	&	1745.0	&	22.997	&	22.698	&	2.6	\\
\hline
\end{tabular}
\end{center}
\label{constants1}
\end{table}

\clearpage

\begin{table}[h]
\caption[]{IR frequencies (cm$^{-1}$) for XH$_3^+$ (X=Ne to
Xe). Italic values correspond to absolute intensities (km/mol) as
obtained at the CCSD/cc-pVQZ level.}
\begin{center}
\begin{tabular}[]{ccccc}
\hline
\hline
\noalign{\smallskip}
& Ne 	& Ar 		& Kr 		& Xe \\
\noalign{\smallskip}
\hline
\noalign{\smallskip}
$\nu_1$	&	3185	  {\it(2)}	&	3311  {\it(25)}	&	3515  {\it(118)}	&	4181  {\it(196)}  \\
$\nu_2$	&	2484  {\it(380)}	&	2104  {\it(106)}	&	1791  {\it(56)}	&	1701  {\it(1941)} \\
$\nu_3$	&	2463  {\it(161)}	& 	1925  {\it(1630)}	& 	1273  {\it(2066)}	 & 804	{\it(3)} \\
$\nu_4$	&	429	{\it(5)	}	& 	803    {\it(2)}	 & 919       {\it(2)} &	579      {\it(11)} \\
$\nu_5$	&	315	{\it(12)}	& 	614    {\it(8)}	& 727     {\it(10)} &	513      {\it(16)}  \\
$\nu_6$	&	314	{\it(417)}	& 	478    {\it(959)}	& 618    {\it(2307)}	& 	 404      {\it(241)} \\
\hline
\end{tabular}
\end{center}
\label{constants2}
\end{table}

\clearpage



\clearpage

\clearpage

\end{document}